# Structural-configurated magnetic plasmon bands in connected ring chains

T. Li[1]*, R. X. Ye[1], C. Li[1], H. Liu[1], S. M. Wang[1], J. X. Cao[1], S. N. Zhu[1]*, and X. Zhang[2]

[1]*National Laboratory of Solid State Microstructures,
Department of Materials Science and Engineering, Department of Physics,
Nanjing University, Nanjing 210093, People's Republic of China*
*taoli@nju.edu.cn, zhusn@nju.edu.cn*

**URL: http://dsl.nju.edu.cn/dslweb/images/plasmonics-MPP.htm**

[2]*5130 Etcheverry Hall, Nanoscale Science and Engineering Center, University of California, Berkeley, California 94720-1740, USA*

**Abstract:** Magnetic resonance coupling between connected split ring resonators (SRRs) and magnetic plasmon (MP) excitations in the connected SRR chains were theoretically studied. By changing the connection configuration, two different coupling behaviors were observed, and therefore two kinds of MP bands were formed in the connected ring chains, accordingly. These MPs were revealed with positive and negative dispersion for the homo- and anti-connected chain, respectively. Notably, these two MP modes both have wide bandwidth due to the conductive coupling. Moreover, the anti-connected chain is found supporting a novel negative propagating wave with a wide band starting from zero frequency, which is a fancy phenomenon in one-dimensional system.

___

## 1. Introduction

Coupled resonator optical waveguide (CROW) was proposed to accommodate the light propagation in a preferred manner due to the coupling between the adjacent resonators [1]. In recent years, using surface plasmon (SP) resonance coupling in arranged metal nanoparticles to constitute CROW has arrested many researchers interest, owing to its ability to confine the energy in a sub-wavelength scale and even able to overcome the diffraction limit [2-4]. On the other hand, artificial magnetic resonator was invented to produce optical magnetic response [5], and it has achieved great progresses in recent years [6-10]. People found such magnetic resonators, such as split ring resonators (SRRs), nano-sandwiches, can be used to construct the CROW as well [11-14], in which a magnetically plasmonic behavior is formed and called as magnetic plasmon (MP). In the work of Ref. [13], conductive coupling based on current exchange in the connected split rings was revealed as a key point to achieve the wide band of MP and low loss, as compared with the former magneto inductive waves [11]. More recently, coupling phenomenon between the artificial magnetic resonances has attracted many attentions [15-18], and the optical stereometamaterials was even achieved associated with the coupling effect [19]. As well as mentioned in Ref. [17] that the coupling between the "magnetic atoms" is a nontrivial mechanism, we would expect diverse functional CROWs with respect to different coupling manner in corresponding to the structural configurations. So, it is undoubted that in a new built CROW system, the coupling mechanism is the most important factor that dominates the characteristic of the waveguide.

In this paper, two kind of structural configurations of SRRs are proposed called homo-connected (slits at same side) and anti-connected (slits at contrary sides), as schematically shown in Fig. 1(b) and (c) respectively. We will firstly make an investigation on the coupling mechanism in these two connected SRR pairs; Different coupling modes are definitely exhibited corresponding to different connection configurations. Afterwards, we extend these SRR pairs to SRR chains by arranging multiple units in one dimension, corresponding to two types of connection configurations as well. As expected, theoretical results show two distinct MP propagation bands with positive and negative dispersions according to different connections. Our study provides flexible method to construct subwavelength waveguides with preferred MP dispersions, and they are expected to be helpful in the development of the Integrated Optics.

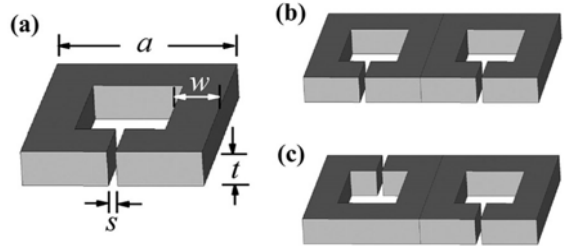

Fig. 1: (a) Scheme of the geometry of single SRR with the parameters marked on. (b) and (c) are the connect SRR pairs with homo- and anti- connection configurations, respectively.

## 2. Resonant coupling between connected SRR pairs

Metallic split ring is a typical artificial magnetic atom proposed by Pendry [5]. Using LC-circuit model, we can describe the SRR's resonant frequency as $\omega_0 = (LC)^{-1/2}$, where $L$ and $C$ are effective inductance and capacitance respectively. Fig. 1(a) shows the square SRR model with the parameters marked on, side length $a$=400 nm, thickness $t$=60 nm, width $w$=100 nm, and slit gap $s$=20 nm. Its electromagnetic (EM) resonance property is numerically evaluated using a commercial software package (CST Microwave Studio), with which the EM response with respect to frequencies and field distributions can be conveniently simulated. Here, the metal is defined as gold, whose permittivity is defined by the Drude model with $\omega_p$=1.37×10$^{16}$ s$^{-1}$ and $\gamma$=4.08×10$^{13}$ s$^{-1}$ [15] (for an ideal model). In order to inspect the response of SRR, we insert a probe inside it (at the center position) with the direction normal to the SRR plane to detect the local magnetic field. In excitations, a discrete port (dipole current) is placed beside the SRR with the current direction parallel to one side. We simulate the model within the frequency ranging from 60 THz to 90 THz. Resonance information are recorded by the probe and field monitors at preferred frequencies. Afterwards, we connect two SRRs with two different configurations, defined as homo- and anti-connections respectively, as shown in Fig. 1(b) and (c). Thus two "magnetic atoms" are bonded together to form a "magnetic molecules" [17], and the same simulation procedure is performed on these two kinds of "molecules".

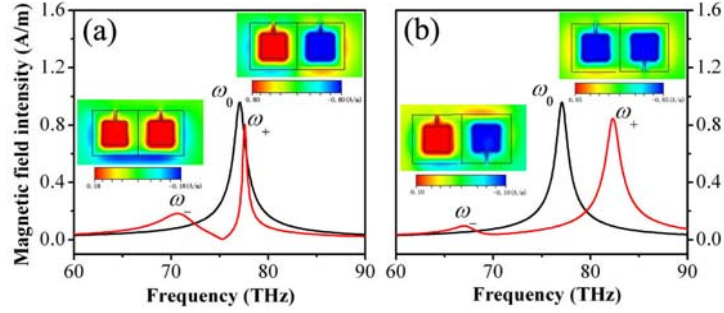

Fig. 2. The magnetic field amplitude intensity detected at the center of the resonators of a single SRR (black curve) and homo-connected SRR pair (red curve) for (a), and anti-connected pair for (b); The insets correspond to the magnetic field maps for the split higher and lower modes, respectively.

Figure 2 shows the simulated results of the resonance properties of single SRR and the coupled cases. As we can see, a single strong peak (black curve) in the detected magnetic field is found at about 77.1THz for the single SRR case, which is definitely an eigen frequency of the LC resonance written as $\omega_0$=77.1 THz. For the coupled cases, two split resonances are clearly observed (red curves) for both homo-connected and anti-connected structures. Although both cases exhibit the split modes due to the coupling effect, the mode shifts are quite different. The split energy of anti-connected case is apparently larger than the other one. From simulations, we get these coupled eigen frequencies as $\omega_{homo+}$=77.6 THz, $\omega_{homo-}$=70.7 THz, $\omega_{anti+}$=82.3 THz and $\omega_{anti-}$=67.0THz, where subscript "+" and "-" correspond to the higher and lower energy levels, respectively. To get a detailed recognition of these coupled modes, magnetic field (perpendicular to the paper) distributions at these frequencies are depicted out as insets in Fig. 2(a) and (b). As expected, these two modes exhibit symmetric and antisymmetric resonance features, as well as the most coupling cases. But what interest us most is that these two coupled structures reverse their eigen modes between higher and lower ones. It is clearly demonstrated that $\omega_+$ corresponds to the antisymmetric mode and $\omega_-$ to the symmetric one for the homo-connected case, and vice versa for the anti-connected case.

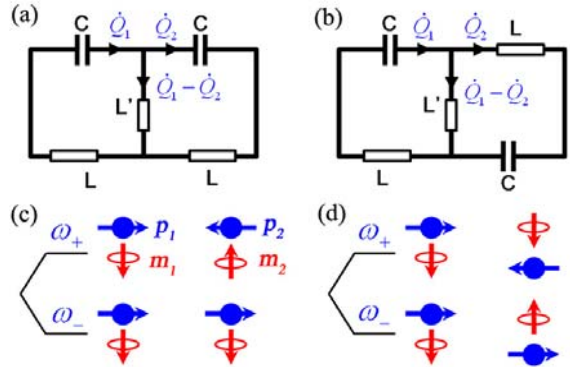

Fig. 3. Equivalent circuits of the two SRR pairs with different connection configurations: (a) homo-connected and (b) anti-connected; Effective configurations of the magnetic and electric dipoles corresponding to the split eigen modes for these two connections, (c)

This result impressively indicates that the coupling behavior of the SRR pairs strongly depends on the connection configurations. To better understand the interactions involved in the splitting of the magnetic resonance, we develop a comparative theoretical analysis based on two coupled LC-circuits. In a common sense, these two kinds of SRR pairs can be regarded as two types of connected LC-circuits, as shown in Fig. 3(a) and 3(b). Here, besides kinetic energy from the inductances $L(\dot{Q}_1^2 + \dot{Q}_2^2)/2$ and the electrostatic energy stored the capacities $(Q_1^2 + Q_2^2)/2C$, the total energy of this system contains three coupling items. The one is the current exchanges [13], and other two is magnetic and electric dipole-dipole interactions, which come from the oscillation moments induced by circuit loops and electric dipoles formed in the capacitances respectively [19]. According the resonance frequency of a single circuit is $\omega_0 = (LC)^{-1/2}$. We can transfer the form of $1/C$ as $L\omega_0^2$. So the Lagrangian of the coupled system is written as

$$\Im = \frac{1}{2}L(\dot{Q}_1^2 + \dot{Q}_2^2) - \frac{1}{2}L\omega_0^2(Q_1^2 + Q_2^2) + \frac{1}{2}L'(\dot{Q}_1 - \dot{Q}_2)^2 - M_m\dot{Q}_1\dot{Q}_2 - M_e\omega_0^2 Q_1 Q_2 \quad (1)$$

where $L'$ is the community inductance from the connected parts, which really contributes a conduction coupling in the SRR pairs; $M_m$ and $M_e$ is the coupling coefficients of the magnetic and electric dipole-dipole interactions, respectively. Here, we do not define the sign of both coefficients just for general evaluations. Actually, they lie on the coupling manners related with the specific structural configurations. Substituting Eq. (1) to the Euler-Lagrangian equations $\frac{d}{dt}(\frac{\partial \Im}{\partial \dot{Q}_i}) - \frac{\partial \Im}{\partial Q_i} = 0, (i=1,2)$, where the Ohmic loss is neglected for a phenomenological explanation, we get two coupled equations,

$$\begin{cases} (L+L')\ddot{Q}_1 + L\omega_0^2 Q_1 - (M_m + L')\ddot{Q}_2 + M_e\omega_0^2 Q_2 = 0 \\ -(M_m + L')\ddot{Q}_1 + M_e\omega_0^2 Q_1 + (L+L')\ddot{Q}_2 + L\omega_0^2 Q_2 = 0 \end{cases} \quad (2)$$

Adopting the root form of $Q_i = A_i \exp(i\omega t)$ and the normalized coupling coefficients as $\frac{L'}{L} = \eta, \frac{M_m}{L} = \kappa_m, \frac{M_e}{L} = \kappa_e$, we can find two eigen-modes as

$$\begin{cases} \omega_1 = \omega_0 \sqrt{\frac{1-\kappa_e}{1+2\eta+\kappa_m}}, & \text{with } Q_1 = -Q_2 \\ \omega_2 = \omega_0 \sqrt{\frac{1+\kappa_e}{1-\kappa_m}}, & \text{with } Q_1 = Q_2 \end{cases} \quad (3)$$

Next, we will get into the specific analysis of the two kinds of SRR pairs, whose split modes are shown in Fig. 2. According the symmetry of the magnetic field of these two modes with respect to the sketched dipole models illustrated in Fig. 3(c) and 3(d), we directly achieve the higher and lower modes with eigen frequency as

$$\omega_+ = \omega_0 \sqrt{\frac{1-\kappa_{e1}}{1+2\eta+\kappa_m}}, \quad \omega_- = \omega_0 \sqrt{\frac{1+\kappa_{e1}}{1-\kappa_m}} \quad , \quad (4a)$$

for the homo-connected case and

$$\omega_+ = \omega_0 \sqrt{\frac{1+\kappa_{e2}}{1-\kappa_m}}, \quad \omega_- = \omega_0 \sqrt{\frac{1-\kappa_{e2}}{1+2\eta+\kappa_m}} \quad , \quad (4b)$$

for the anti-connected one. Here, $\kappa_{e1}$ and $\kappa_{e2}$ correspond to the electric coupling coefficients of the two different connection cases. Since we have already obtained the eigen frequencies of the two coupled cases from the simulations, coupling coefficients are easily to be calculated out as $\eta=$ 0.072, $\kappa_m=$ 0.007, $\kappa_{e1}=$-0.166 and $\kappa_{e2}=$ 0.131. These data, therefore, provides us a ruler to evaluate the coupling strength as we are stepping into the underlying physics of this coupled system. As we know that the energy of the dipole-dipole interaction (both electric and magnetic) has the form of

$$\mathbf{E}_{dipole} = \frac{\mathbf{d}_1 \cdot \mathbf{d}_2}{r^3} - \frac{3(\mathbf{d}_1 \cdot \mathbf{r})(\mathbf{d}_2 \cdot \mathbf{r})}{r^5} \quad , \quad (5)$$

where $\mathbf{d}_i$ (i=1, 2) represent either the magnetic dipoles ($\mathbf{m}_i = \dot{Q}_i S$) or electric dipoles ($\mathbf{p}_i = Q_i l$).

Comparing the equivalent coupled LC-circuits and effective dipoles configurations of split modes (see Fig. 3), we get easy to draw the conclusion that the coefficients of magnetic coupling of the two cases are both positive, while that of electric coupling should be negative for homo-connect SRR pair and positive for the anti-connected one. Regarding the different distance and relative orientations of the electric dipoles ($\mathbf{p_1}$, $\mathbf{p_2}$), the absolute value of $\kappa_{e1}$ should be evidently larger than $\kappa_{e2}$, which agrees well with retrieved results through the former equations.

Among the interaction items, we also find that the electric coupling ($\kappa_e$) is the strongest one while the magneto induction ($\kappa_m$) is much weaker. Thus, alternating the position of the slit changes the contribution of electroinductive coupling and therefore reverse split eigen modes, which appropriately explains the simulation results. Furthermore, it tells us that although the system manifests a feature of magnetic resonance, the major coupling components are from the electric dipole-dipole interaction and current exchange. Here, the current exchange contributing the conductive coupling is accommodated in the community inductance, which is expected playing an important role to build a wide MP band as we extents these SRR pairs to the SRR chains exhibited in the next section.

## 3. MP modes in the chains of connected SRRs

Inspired by above results, we will consciously prospect the circumstances of the long SRR chains by these two kinds of connections. Due to confinement in the transverse dimension, such chains are regarded as subwavelength waveguides. They are another kind of CROWs based on the coupling between artificial magnetic resonances. Thus, we consequently construct two CROWs made of the connected SRR chains both containing 80 units, which are homo- and anti-connected as schematically shown in Fig. 4(a) and 4(b), respectively. Numerical simulations are still performed using the commercial solver (CST Microwave Studio). An about 0.01ps pulse from a dipole source beside the left side of the chain was defined as the excitation signal, which covers a wide range from 0 to 100THz in the frequency domain. After numerical calculations modeled with the open boundary condition, the solver provides us the whole field distributions at monitored frequencies (from 0 to 100THz). Afterwards, we extract the normal component (perpendicular to SRR plane) of magnetic field in the center line at a serial frequencies from 0 to 100THz, and use a Fourier transformation (FT) method to transform them from spatial region to wave vector region [14],

$$H(\omega, k) = \int H(\omega, x) e^{ikx} dx . \tag{6}$$

By this means, magnetic field distribution in the $\omega$-$k$ space can be obtained, which may give a clear picture of the dispersion property of the guided wave via the coupled resonances.

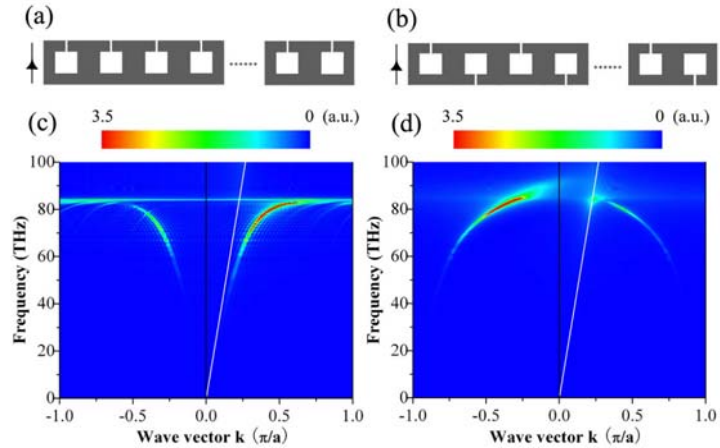

Fig.4 Subwavelength waveguides constituted by the SRR chains with (a) homo-connection and (b) amti-connection. (c) and (d) are the calculated corresponding magnetic field map in the $\omega$-$k$ space with respect to the waveguides (a) and (b) respectively, in which dispersions of MP waves propagating with the waveguides are manifested clearly.

From the simulated results, we find that at some frequencies (~80 THz) the dipole excited field can propagate very far and does not tail off even at the 80th unit for both SRR chains. They actually exhibit the characteristics of a good subwavelength waveguide. To get the most important features of waveguide properties of these two CROWs, we would direct into the dispersions revealed the H field map in $\omega$-$k$ space. Firstly for the homo-connected SRR chain shown in Fig 4(c), a clear dispersion similar to the Surface Plasmon (SP) in flat metal surface is observed, together with a marked light line in free space that totally lies above the MP mode. This dispersion exhibits two important aspects. The one is the coupled SRR chain merges the discrete modes into a wide continuous band even extending to the zero-frequency (although very weak), which is never realized in nano particles or sandwich chains [12, 14]. The other is that the dispersion curve meets an upper limit at about 82 THz, where it tends to be flat and the supported wave should have very slow group velocity ($v_g=\partial\omega/\partial k$) and very large density of state (DOS). This upper limit frequency is very like the well-known Surface Plasmon frequency ($\omega_{sp}$), so as to be characterized as $\omega_{mp}$, which undoubtedly can be modulated by the structural parameters of SRR chain. Moreover, compared with Ref. [13], FT method allows us to achieving a whole dispersion map of the MP wave covering the negative and positive $k$ regions with the intensity resolution. It should be mentioned that the dispersion map are extracted directly from the field distributions for a specific excitation case that the source is placed at the left side. Therefore, the dispersion curve with positive $k$ is well anticipated as has been displayed in Fig. 4(c). However, the negative propagation MP wave with negative $k$ is also presented symmetrically to the positive branch. After a little analysis, we may be aware of that it is actually a reflection wave that comes from the propagating wave reflected by every SRR units, and it is reasonably weaker than the positive direction wave.

As for the anti-connected SRR chain, a fancy result appears. As well as the reversed eigen modes in the anti-connected SRR pair, an obvious negative dispersion with a wide band is revealed for the MP wave in this long chain as shown in Fig. 4(d). Different from the former one, however, it meets a upper limit at the Brillion center and extends to zero frequency at the Brillion boundary that rightly reverse the dispersion of homo-connected case. Moreover, we notice that the field intensity at negative $k$ region is much stronger than that in the positive one. That means the MP propagation in this anti-connected SRR chain is still a front-propagating wave (energy flows along right direction), with a little reflection backward. However, the negative dispersion results in an anti-paralleled group and phase velocity, $(\omega/k)(\partial\omega/\partial k)<0$, this propagating wave thus can be regarded as a negative refraction. It is a significant finding because this negative refraction wave revealed here has a wide bandwidth even extending to zero frequency, which was never reported in the one-dimensional system. In addition, a cross point of the MP wave with light in free space is clearly observed in positive $k$ region of the dispersion map [$k$~0.23 ($\pi$/a) and $\omega$~84 THz], which can be called as "Magnetic Plasmon Polariton-MPP". This MPP mode is regarded as a phase matching point as well. Since the light radiation from the dipole source only propagates from left to right [see the sketch Fig. 4(a) and (b)] with a positive $k$, there is no MPP appear in the negative $k$ region.

Now, we have numerically revealed MP wave propagation behavior in the connected SRR chain composed CROWs with two different connection configurations. The reversed dispersion characteristics also can be explained by extending the coupled LC-circuit theory of the SRR pairs elucidated in section 2 to a chain system, whereas it may be more complicated. Here, we only want to focus our attention at the physics behind the simulations instead of to be involved in the massive mathematical procedures. Referring to the items in the Lagrangian of connected SRR pairs [Eq. (1)], we can find that except the self energy of kinetic item of $L$ and potential of $C$, the coupling include three items, (a) the community current exchanges in connected segments, (b) magnetic dipole-dipole interaction, and (c) electric dipole-dipole interaction. As well as the connected SRR pairs as has been elaborately discussed in section 2, the reversal of the dispersion is mainly come from the alternation of the electroinductive coupling due the change of the slits configuration. Like the case of Ref. [13], the conductive item (a) attributing from the current exchanges is an important factor to build such a wide MP band, which do not exist in the coupling between the nanoparticles, nanosandwiches, or some other discrete resonators [2-4, 12, 18-19]. So the MP waveguides proposed here reveal a real plasmonic behavior very similar to the SP in a flat metallic surface. At this point, our study provides another method to construct subwavelength CROWs with wide band that accommodating the MP wave propagation with in a preferred characteristics. And the Magnetic Plasmon frequency ($\omega_{mp}$) is a solely controllable character that can be modulated by the structural parameters. More significantly, the anti-connected SRR chain exhibits a fancy negative refractive property, whose phase and group velocities are anti-paralleled. This negative dispersion MP wave is demonstrated by the simulated movie as well (see the supplemental material online).

**4. Conclusions**

In summary, we have investigated the coupling mechanism in two types of connected SRR pairs, as well as the magnetic plasmon (MP) modes in long SRR chains. Three factors of the coupling, current exchange, magnetic dipolar and electric dipolar interactions, are emphatically studied. Numerical simulation and analytical deducing indicate that electroinductive coupling plays an important role in the coupling of connected SRRs and even reverse the modes for different connections. As for the long CROWs constructed by the SRR chains, the homo-connected one supports a SP-like positive directional MP wave while the anti-connected one supports a negative directional one. Thanks to the conductive coupling from the current exchanges in this system, both MP modes exhibit extraordinary wide band extending to zero frequency. The artificial controllability of the MP modes would enable such kind of subwavelength CROWs to be a promising candidate for the development of Integrate Optics or some other nano-optical devices.

**Acknowledgments**

This work was supported by the State Key Program for Basic Research of China (Nos. 2006CB921804 and 2009CB930501), the National Natural Science Foundation of China under Contract Nos. 10704036, 10604029, 10534042, and National Fundamental Fund of Personnel Training (No. J0630316).